\documentclass[11pt]{article}
\usepackage{acl}
\usepackage{times}
\usepackage{latexsym}
\usepackage[T1]{fontenc}
\usepackage[utf8]{inputenc}
\usepackage{microtype}
\usepackage{inconsolata}
\usepackage{graphicx}
\usepackage{booktabs}
\usepackage{multirow}
\usepackage{amsmath}
\usepackage{enumitem}
\usepackage{listings}
\usepackage{xcolor}
\usepackage{subcaption}
\usepackage{url}
\usepackage{placeins}

\title{RepoNav: From Snippet Retrieval to File-Centered Repository Navigation for Code Agents
}

\author{
Hongzheng Chai$^{1}$ \quad
Jiakun Li$^{1}$ \quad
Hongyue Yu$^{2}$ \quad
Yuan Yuan$^{1,3,4}$\thanks{Corresponding author.} \\
$^{1}$School of Computer Science and Engineering, Beihang University \\
$^{2}$National College for Excellent Engineers, Beihang University \\
$^{3}$Hangzhou Innovation Institute, Beihang University \\
$^{4}$Qingdao Research Institute, Beihang University \\
\texttt{chaihongzheng@buaa.edu.cn, yuan21@buaa.edu.cn}
}

\begin{document}
\maketitle

\begin{abstract}
Solving repository-level code tasks requires LLM-based agents to use code search tools to navigate large codebases and identify a small set of relevant files and functions.
However, current retrieval tools typically return flat lists of isolated code snippets: such lists can surface relevant files, but provide insufficient structure for agents to distinguish the target function from semantically similar alternatives in the same file.
We introduce RepoNav, a lightweight post-retrieval interface that reorganizes retrieved snippets into a file-centered navigation scaffold.
By presenting compact structural cues and candidate targets, this scaffold guides on-demand file-structure browsing, helping agents compare sibling symbols before selecting a target function.
Across diverse models on LocBench, RepoNav improves function-level localization and narrows the file-to-function gap. Controlled ablations demonstrate that these gains come from structured evidence organization rather than simply exposing additional file structure, and the approach also improves performance on a repository-level question-answering benchmark.
\footnote{We release code at \url{https://github.com/CCChz233/reponav} to facilitate future research.}
\end{abstract}

\section{Introduction}

Large language models (LLMs) have driven rapid progress in autonomous software engineering.
In repository-scale tasks such as issue resolution and bug localization, agents must operate across two levels of granularity: identifying the relevant files and pinpointing the specific functions within them.
Modern code agents equipped with retrieval or agentic search tools~\citep{repocoder,CodeRAG-Bench,sweagent} can often surface relevant files, yet function-level localization still lags significantly behind.
We analyze this behavior using a common mini-SWE-agent scaffold~\citep{sweagent}, keeping the underlying agent loop fixed across exploration interfaces.

\begin{figure}[tb]
    \centering
    \includegraphics[width=\columnwidth]{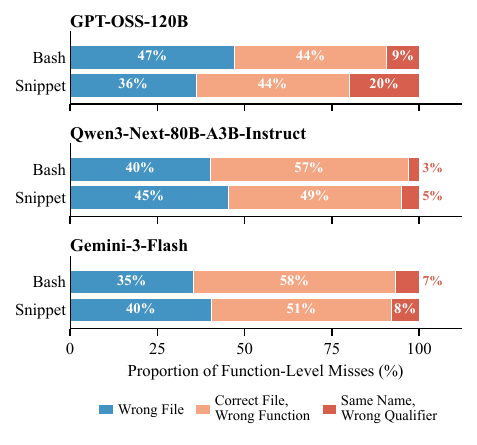}
    \vspace{-1.5em}
    \caption{Failure taxonomy for function-level misses from interactive agent executions.
Percentages are computed over misses within each LLM backbone and setting.
A substantial fraction of misses occurs after the agent has reached the correct file, indicating within-file navigation failures.
Percentages may not sum to 100 due to rounding.}
    \label{fig:failure_taxonomy}
     \vspace{-0.5em}
\end{figure}

Why do agents that reach the correct file still miss the target function?
Our analysis reveals \emph{premature anchoring}: agents often commit to a salient nearby symbol before inspecting sibling definitions or file-structure views.
Flat snippet retrieval reinforces this behavior by presenting relevant chunks and distractors as isolated evidence, with limited cues about what else the file contains.
As Figure~\ref{fig:failure_taxonomy} shows, \emph{Correct File, Wrong Function} accounts for 44--58\% of function-level misses across agents instantiated with different LLM backbones, confirming within-file navigation as a central bottleneck in function-level localization.

To mitigate this bottleneck under a fixed retrieval substrate, we introduce RepoNav, a lightweight post-retrieval interface for repository navigation.
Here, \emph{lightweight} refers to low incremental infrastructure and deployment overhead: RepoNav reuses the existing dense retrieval substrate and requires no additional persistent structural index or repository-wide graph.
RepoNav reorganizes the retrieved evidence into a file-centered navigation scaffold.
The scaffold exposes compact structural cues, candidate targets, and continuation hints, and supports on-demand file-structure browsing through \texttt{list\_symbols}.
This design separates file discovery from within-file verification: retrieval proposes candidate files, while RepoNav helps the agent decide which internal symbols to inspect next.
Throughout this paper, we use \emph{symbol} to denote a named structural code unit, such as a function, method, or class.
Rather than asking agents to read more code upfront, RepoNav makes candidate files structurally expandable and encourages verification of sibling symbols before commitment.
Unlike heavyweight graph-based systems~\citep{Repograph,Codexgraph,locagent} that rely on repository-wide graph construction or static-analysis infrastructure, RepoNav is an agent-facing interface layer built on top of existing snippet retrieval outputs.

Across seven models on LocBench~\citep{locagent}, RepoNav improves function-level localization and reduces the gap between file discovery and function discovery.
A tool-matched Snippet+ListSym baseline shows that file-structure browsing is useful, while RepoNav further improves over this baseline by making such browsing more actionable through file-centered organization.
Results on SWE-QA-Bench~\citep{Peng2025SWEQACL} show that RepoNav also improves repository-level question answering beyond explicit localization.
Our contributions are:
\begin{enumerate}[leftmargin=1.5em,itemsep=2pt,parsep=0pt]
    \item \textbf{An empirical diagnosis of the file-to-function gap.}
    We show that a large fraction of function-level failures occur after the agent has already reached the correct file, identifying within-file navigation as a major bottleneck in repository-scale code localization.

    \item \textbf{A structurally expandable, file-centered navigation interface.}
    RepoNav reorganizes flat snippet-style outputs into file-centered scaffolds while keeping the underlying dense index and raw retrieval scores fixed.
    It integrates \texttt{list\_symbols} as an on-demand file-structure browsing tool, enabling agents to inspect lightweight file skeletons and compare sibling symbols without reading full file bodies.

    \item \textbf{Behavioral and tool-matched evidence for scaffolded navigation.}
    A Snippet+ListSym baseline controls for access to file-structure browsing, and controlled ablations show that RepoNav provides additional gains by making structural evidence more actionable and efficient for agent exploration.
\end{enumerate}

\begin{figure*}[!t]
    \centering
    \includegraphics[width=0.95\textwidth]{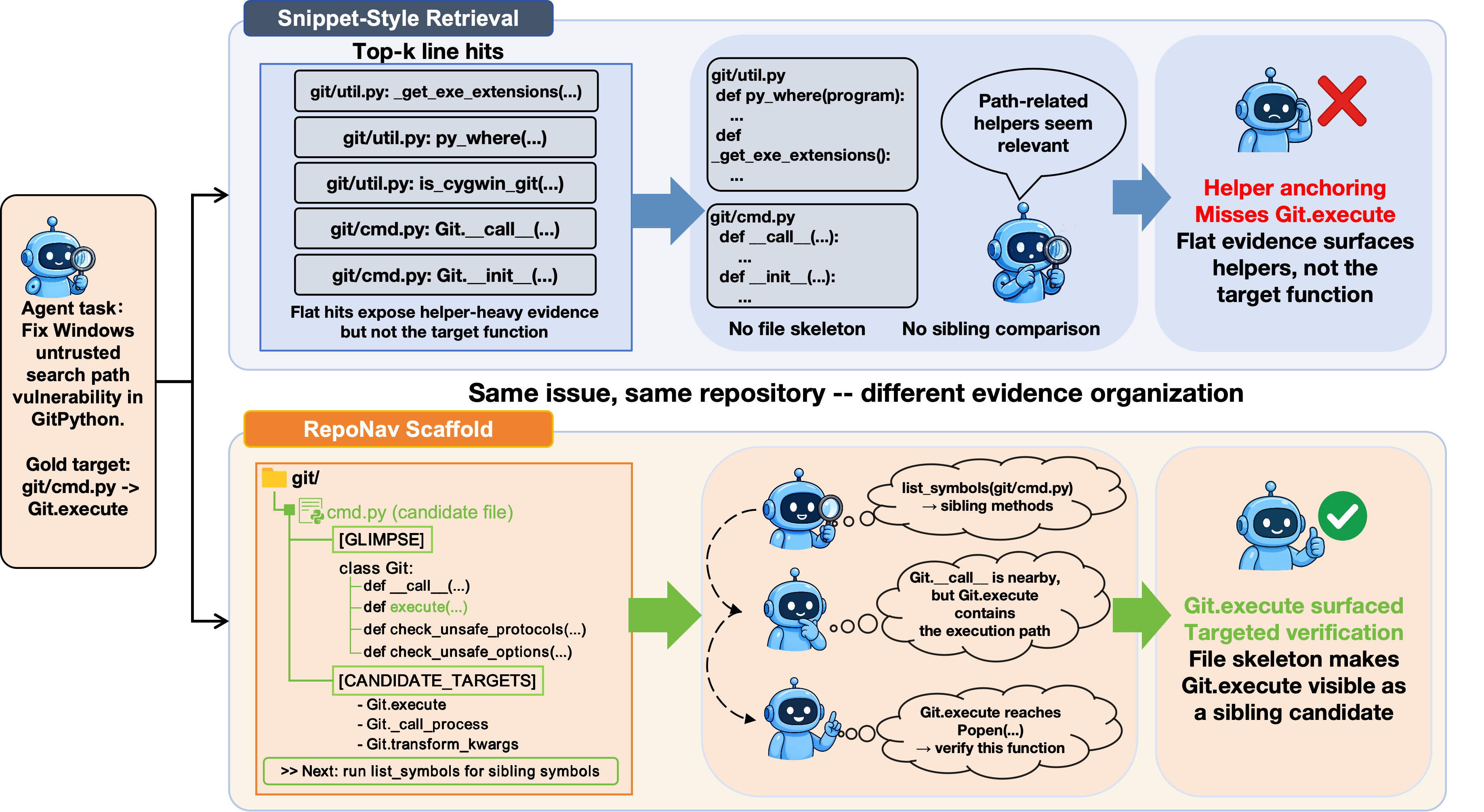}
  \caption{
A real GitPython case illustrating function-level divergence under different evidence organization.
Top: snippet-style retrieval returns helper-heavy flat evidence and misses \texttt{Git.execute}.
Bottom: RepoNav organizes the candidate file into a file-centered scaffold and prompts on-demand file-structure browsing with \texttt{list\_symbols}, making \texttt{Git.execute} visible as a sibling candidate for targeted verification.
}
    \label{fig:motivating_example}
\end{figure*}

\section{Background and Related Work}
\label{sec:motivation}

\subsection{Code Agents and Exploration Interfaces}
\label{sec:evidence-action-gap}

SWE-bench~\citep{swebench} has become a standard benchmark for autonomous code agents~\citep{AutoCodeRover,sweagent,OpenHands,MoatlessTools,Agentless},
which navigate repositories through interleaved reasoning and tool use~\citep{react}.
These agents typically interact with repositories through bash-style exploration or snippet-style search interfaces.
Retrieval-augmented generation~\citep{RAG} has been widely adopted for injecting external knowledge into language models; in the code domain, retrieval-based methods such as RepoCoder~\citep{repocoder} and CodeRAG-Bench~\citep{CodeRAG-Bench} extend this paradigm to repository-level tasks.
However, snippet-search interfaces usually present retrieved chunks as flat, chunk-centered lists, while bash-style interfaces expose the raw repository but require the agent to infer structure from command outputs.
In both cases, the interface provides limited support for turning retrieved evidence into navigable action choices.
This limitation is especially problematic for function-level localization, where the agent must compare multiple candidate symbols after reaching a relevant file rather than merely identify the file itself.

\subsection{The File-to-Function Gap}
\label{sec:file-vs-function}

A natural assumption is that function-level localization failures mainly arise because the correct file was never surfaced.
Our failure analysis shows otherwise.
We categorize each function-level miss into three types:
\emph{Wrong File} (the predicted target lies outside the gold file),
\emph{Correct File, Wrong Function} (the agent reaches the correct file but selects the wrong function), and
\emph{Same Name, Wrong Qualifier} (the prediction matches the local symbol name but not the fully qualified target).
As Figure~\ref{fig:failure_taxonomy} shows, a substantial share of misses occurs after the agent has already reached the correct file: \emph{Correct File, Wrong Function} accounts for 44--58\% of all misses.
This reveals a file-to-function localization gap: reaching the correct file does not reliably translate into identifying the correct function.

\paragraph{Premature anchoring.}
\label{sec:premature-anchoring}
This asymmetry gives rise to a recurring failure mode: the agent commits to the most salient nearby symbol before inspecting sibling or structurally adjacent candidates.
Typical anchors include public wrappers, request handlers, or entry methods, as illustrated in Figure~\ref{fig:motivating_example} (see Appendix~\ref{sec:appendix-cases} for further case studies).
This behavior is reminiscent of the ``lost in the middle'' phenomenon in long-context models~\citep{liu-etal-2024-lost}, where information position affects utilization; here, the agent fixates on the most prominent symbol rather than systematically comparing alternatives.
The interface does not encourage continued exploration, motivating a navigation-oriented view: post-retrieval interfaces should organize evidence into actionable choices that make continued exploration easier than premature commitment.

\paragraph{Positioning.}
Recent efforts have scaled structural representations to the repository level.
RepoGraph~\citep{Repograph} and CodexGraph~\citep{Codexgraph} map entire codebases into graph databases; LocAgent~\citep{locagent} constructs directed code graphs for multi-hop traversal; LingmaAgent~\citep{LingmaAgent} builds repository-level
knowledge representations with Monte Carlo tree search for exploration; and GraphCoder~\citep{GraphCoder} uses code context graphs for retrieval-augmented completion.
These methods rely on repository-wide graph construction and specialized traversal or query mechanisms, which introduce additional infrastructure and interaction complexity for agent systems.
In the broader NLP setting, StructRAG~\citep{StructRAG} converts retrieved documents into task-appropriate structured formats to aid reasoning; RepoNav shares this motivation of post-retrieval restructuring but targets code agents and operates as a lightweight interface layer rather than a general document transformation framework.
Rather than precomputing a global graph, RepoNav dynamically extracts local structural cues from only the top-ranked retrieved files and presents them as an immediately consumable text-based scaffold.

\paragraph{Research questions.}
The preceding analysis motivates three questions:
(\textbf{RQ1})~Does the file-to-function gap exist consistently across models, and does a navigation-oriented interface reduce it?
(\textbf{RQ2})~Is the behavioral shift driven by the \emph{volume} of structural information or by how that information is \emph{organized}?
(\textbf{RQ3})~Does the benefit transfer beyond localization to repository-level question answering?

\section{RepoNav: A Navigation Interface for Repository Exploration}

RepoNav is a post-retrieval interface layer rather than a new retriever.
It does not modify the raw dense chunk retrieval results or similarity scores used by the snippet-search baseline.
Instead, it deterministically aggregates retrieved chunks into file-level candidates and changes how the retrieved evidence is presented to the agent.
RepoNav turns raw retrieval hits into cues for comparing plausible symbols before commitment.

\subsection{Interface Design}
\label{sec:design-principles}

RepoNav follows three design principles.
First, it aggregates chunk-level evidence into candidate files, since agents ultimately inspect, reason over, or modify files rather than isolated chunks.
Second, it exposes compact file-internal structure, giving the agent enough context to compare nearby symbols without dumping the full file.
Third, it adds explicit continuation cues, so that retrieved evidence is treated as a starting point for targeted inspection rather than as a terminal answer.

Concretely, RepoNav serializes the top-ranked files as a plain-text, indentation-based tree.
Each file entry contains three blocks.

\paragraph{\texttt{[ANCHORS]}: Entry points.}
After dense retrieval results are aggregated into candidate files, RepoNav identifies \emph{anchor symbols} within those files using a lightweight lexical match between symbol names and the retrieval query issued by the agent.
These anchors provide query-relevant entry points into each file.
When no symbol name matches the query, RepoNav falls back on the structural sketch described below.
Each anchor is annotated with its line span and compact same-file call context, giving the agent a grounded starting point for inspection.
In our implementation, anchors are capped at two per file.

\paragraph{\texttt{[GLIMPSE]}: Structural sketch.}
To prevent fixation on anchors, RepoNav exposes up to three non-anchor symbols from the same file, or up to five when the top-ranked file has no anchor match.
These symbols provide a compact sketch of what else the file contains, such as additional classes, functions, or methods that may not directly match the query but are structurally relevant.
Rather than dumping full function bodies or long signatures, \texttt{[GLIMPSE]} presents schematic entries, allowing the agent to notice alternative candidate symbols with minimal context cost.

\paragraph{\texttt{[CANDIDATE\_TARGETS]}: Actionable target shortlist.}
This block turns the structural sketch into a compact set of candidate inspection targets.
It summarizes anchors, selected glimpse symbols, and lightweight same-file call context without adding new retrieval evidence or full function bodies.
The shortlist is capped at four entries and paired with fixed continuation hints, encouraging the agent to inspect sibling symbols before committing.
The call context serves only as a navigation cue rather than sound static analysis.

\paragraph{\texttt{list\_symbols}: File-structure browsing.}
RepoNav exposes \texttt{list\_symbols} as an on-demand file-structure browsing tool.
Given a file path, it returns a lightweight file skeleton, including imports, classes, functions, methods, line ranges, and optional signatures, but not full function bodies.
It does not perform semantic search or rank target functions.
Instead, it lets the agent inspect the editable objects inside a candidate file and compare sibling symbols before selecting a target function.

The scaffold itself remains plain text and requires no graph query language.
It can therefore be consumed by standard text-based code agents while still exposing enough structure to support targeted navigation.
The full serialization budgets, ordering rules, and a worked output example are provided in Appendix~\ref{app:scaffold-spec} (Figure~\ref{fig:reponav_output}).

\subsection{Implementation}
\label{sec:implementation}

RepoNav is implemented as a two-stage post-retrieval pipeline.
Both stages require minimal computation and no offline preprocessing beyond the standard dense chunk index used by the snippet baseline.
The specific retrieval model and indexing configuration are detailed in Section~\ref{sec:setup}.

\paragraph{Stage 1: Evidence aggregation.}
Given an agent retrieval query, the dense retrieval backend returns top-ranked code chunks and their associated retrieval scores.
RepoNav aggregates these chunk-level scores into a file-level ranking using a hybrid scoring rule.
For each candidate file $f$, let $C_f$ denote the retrieved chunks from $f$, with their similarity scores sorted as
$s_{f,(1)} \ge s_{f,(2)} \ge \dots$.
We aggregate the top $m_f=\min(m, |C_f|)$ scores as:
\[
\mathrm{Score}(f)
=
\alpha s_{f,(1)}
+
(1-\alpha)\frac{1}{m_f}\sum_{i=1}^{m_f}s_{f,(i)} .
\]
We set $m{=}3$ in all experiments.
Files with fewer than $m$ retrieved chunks are averaged over their available chunks.
The first term captures peak evidence strength, while the second rewards files supported by multiple high-scoring regions rather than a single spurious match.
A parameter sensitivity analysis in Appendix~\ref{sec:appendix-sensitivity} shows that performance is stable across a wide range of $\alpha$ values, with ranking-oriented metrics peaking near $\alpha{=}0.5$ and recall metrics plateauing for $\alpha \ge 0.7$.
Since $\alpha$ only affects the file ranking before agent interaction, we select its value on held-out development splits using offline retrieval metrics and then keep it fixed for all downstream experiments.

\paragraph{Stage 2: On-the-fly structural extraction.}
RepoNav dynamically parses only the top-$k$ files from Stage~1, with $k{=}5$ fixed across experiments by default.
The initial scaffold auto-expands full three-block entries for the top three candidate files, while the remaining parsed files remain available for on-demand structure browsing.
For Python files, the built-in \texttt{ast} module extracts function and class definitions, import statements, and a lightweight same-file call neighborhood.
The call neighborhood connects functions only when a call expression can be matched to another function defined in the same file; it does not attempt sound static analysis, cross-file dependency recovery, or interprocedural call-graph construction.
When call information is unavailable, RepoNav falls back to file-structure skeletons without caller and callee cues.
These extracted cues are assembled into the three-block interface described in Section~\ref{sec:design-principles}.

\begin{figure}[htb]
  \centering
  \includegraphics[width=\columnwidth]{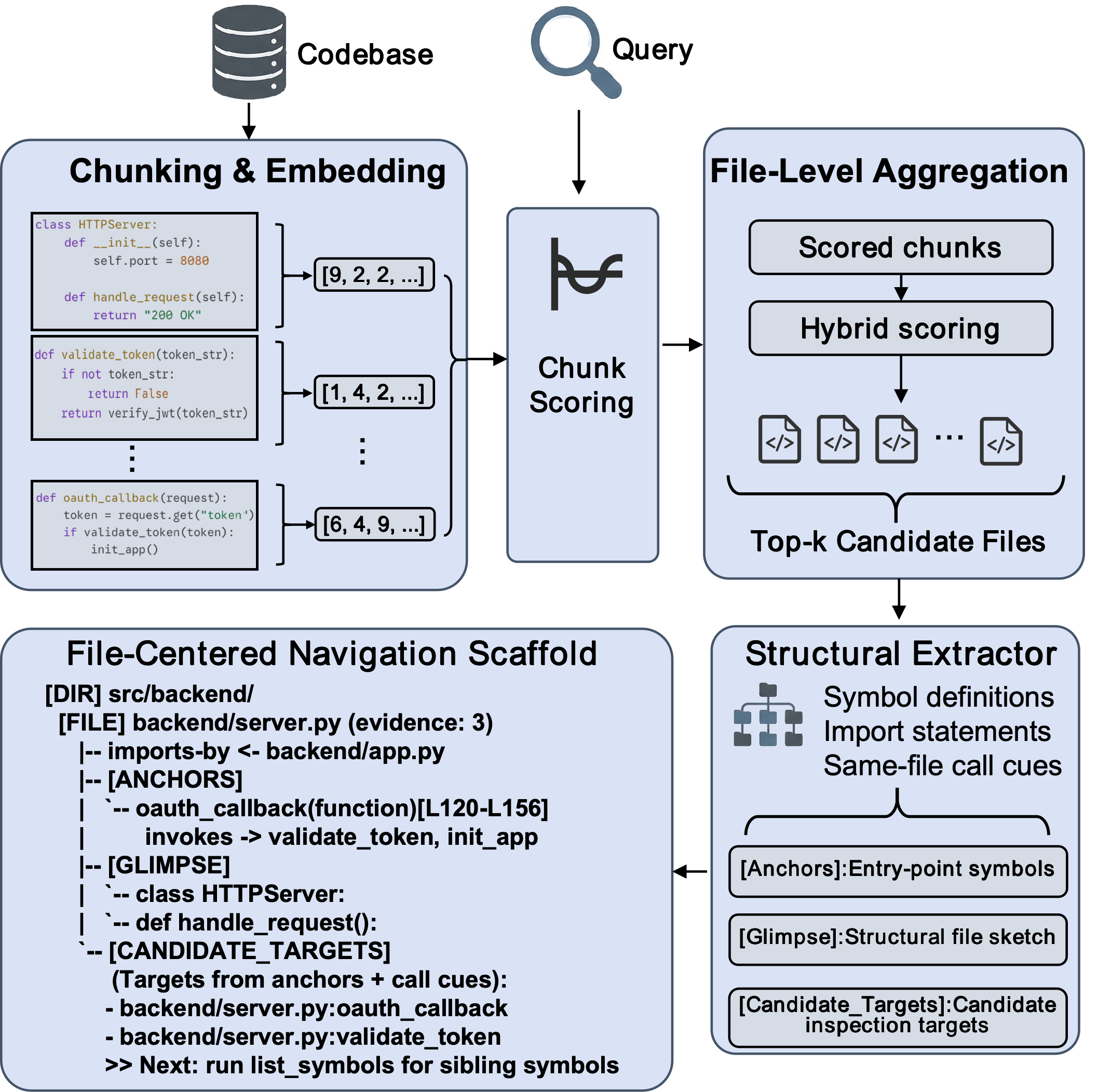}
  \caption{
  Implementation overview of RepoNav.
  The retrieval backend is shared with the snippet-search baseline; RepoNav changes the post-retrieval organization of evidence by aggregating chunks into candidate files and extracting lightweight file structure.
  }
  \label{fig:pipeline}
  \vspace{-0.5em}
\end{figure}

Figure~\ref{fig:pipeline} summarizes the two-stage post-retrieval pipeline.
The process is deterministic and local: RepoNav aggregates retrieved chunks into candidate files and extracts structure only from the top-ranked files, avoiding repository-wide graph construction.
This design allows evidence from multiple retrieved regions to jointly support a file while retaining multiple candidate files for subsequent within-file inspection and cross-file pivoting.

\section{Experimental Evaluation}

We now evaluate the three research questions introduced in Section~\ref{sec:motivation}: whether the file-to-function localization gap holds consistently across models and can be reduced by a navigation-oriented interface (RQ1), whether the observed behavioral shift is driven by the volume of structural information or by how that information is organized (RQ2), and whether the benefit transfers beyond localization to repository-level question answering (RQ3).

\begin{table*}[tb]
\centering

\scriptsize
\setlength{\tabcolsep}{2.3pt}
\resizebox{\textwidth}{!}{%
\begin{tabular}{lcccccccccccccccc}
\toprule
& \multicolumn{4}{c}{File Acc@5}
& \multicolumn{4}{c}{Module Acc@5}
& \multicolumn{4}{c}{Function Acc@5}
& \multicolumn{4}{c}{Function Rec@10} \\
\cmidrule(lr){2-5}
\cmidrule(lr){6-9}
\cmidrule(lr){10-13}
\cmidrule(lr){14-17}
Model
& B & S & S+L & R
& B & S & S+L & R
& B & S & S+L & R
& B & S & S+L & R \\
\midrule
Qwen2.5-72B
& 0.589 & 0.680 & 0.673 & \textbf{0.698}
& 0.396 & 0.461 & 0.562 & \textbf{0.570}
& 0.254 & 0.300 & 0.384 & \textbf{0.434}
& 0.307 & 0.377 & 0.490 & \textbf{0.549} \\

GPT-OSS-120B
& 0.705 & 0.759 & 0.759 & \textbf{0.770}
& 0.500 & 0.613 & 0.630 & \textbf{0.650}
& 0.339 & 0.464 & 0.488 & \textbf{0.527}
& 0.417 & 0.571 & 0.586 & \textbf{0.618} \\

Qwen3-Next-80B
& 0.734 & 0.732 & 0.733 & \textbf{0.739}
& 0.523 & 0.568 & 0.589 & \textbf{0.632}
& 0.339 & 0.411 & 0.454 & \textbf{0.498}
& 0.410 & 0.517 & 0.564 & \textbf{0.622} \\

Qwen3-Coder-30B
& 0.700 & 0.693 & 0.707 & \textbf{0.727}
& 0.564 & 0.584 & 0.598 & \textbf{0.629}
& 0.420 & 0.459 & 0.476 & \textbf{0.516}
& 0.498 & 0.551 & 0.572 & \textbf{0.618} \\

MiniMax-M2.5
& 0.760 & 0.764 & 0.780 & \textbf{0.803}
& 0.660 & 0.671 & 0.671 & \textbf{0.691}
& 0.525 & 0.532 & 0.534 & \textbf{0.601}
& 0.633 & 0.641 & 0.652 & \textbf{0.708} \\

GLM-4.7
& 0.808 & 0.821 & 0.812 & \textbf{0.839}
& 0.685 & 0.698 & 0.709 & \textbf{0.734}
& 0.585 & 0.600 & 0.577 & \textbf{0.632}
& 0.714 & 0.722 & 0.695 & \textbf{0.783} \\

Gemini-3-Flash
& 0.836 & 0.823 & 0.822 & \textbf{0.840}
& 0.709 & 0.702 & 0.717 & \textbf{0.752}
& 0.625 & 0.623 & 0.648 & \textbf{0.689}
& 0.737 & 0.722 & 0.739 & \textbf{0.803} \\
\midrule
Average
& 0.733 & 0.753 & 0.755 & \textbf{0.774}
& 0.577 & 0.614 & 0.639 & \textbf{0.665}
& 0.441 & 0.484 & 0.509 & \textbf{0.557}
& 0.531 & 0.586 & 0.614 & \textbf{0.672} \\
\bottomrule
\end{tabular}%
}
\caption{
Main LocBench localization results across seven models.
We report File Acc@5, Module Acc@5, Function Acc@5, and Function Rec@10.
Methods are abbreviated as \textbf{B} = Bash, \textbf{S} = Snippet Search, \textbf{S+L} = Snippet+ListSym, and \textbf{R} = RepoNav; \textbf{S+L} is a tool-matched snippet baseline with the same \texttt{list\_symbols} access as RepoNav.
Bold marks the best setting within each model block and metric group.
Full Accuracy@$k$ and Recall@$k$ results are provided in Appendix~\ref{sec:appendix-full}, Table~\ref{tab:full-results}.
}
\label{tab:locbench-main}
\end{table*}

\subsection{Experimental Setup}
\label{sec:setup}

\paragraph{Benchmark.}
We evaluate on LocBench~\citep{locagent}, a repository-level code localization benchmark with 560 Python instances, each pairing a natural language issue with gold files and functions from the corresponding patch.

\paragraph{Agent settings.}
All experiments use a common mini-SWE-agent scaffold~\citep{sweagent} with a fixed agent loop, environment, and interaction budget.
We compare four interfaces:
\textbf{Bash} (shell only),
\textbf{Snippet Search} (adding a dense search tool returning flat chunk lists),
\textbf{Snippet+ListSym} (adding the same \texttt{list\_symbols} tool used by RepoNav to the flat snippet interface, controlling for file-structure browsing access), and
\textbf{RepoNav} (reorganizing the same dense retrieval substrate into a file-centered scaffold with \texttt{list\_symbols} access).

\paragraph{Retrieval settings.}
All three retrieval-based settings share the same chunk index, embedding model, retrieval backend, and raw chunk-level similarity scores.
Following \citet{locagent}, each function is embedded as a single chunk using CodeRankEmbed~\citep{CoRNStack}.
The main experiments retrieve 80 raw chunks per query before file-level aggregation.
CodeRankEmbed is the primary retriever used in the full experiments.
We additionally conduct a fixed 100-instance pilot with UniXcoder~\citep{guo-etal-2022-unixcoder} to examine compatibility with a second embedding model.
Retrieval-depth sensitivity and alternative-embedder results are reported in Appendix~\ref{app:retrieval-details}.

\paragraph{Models and metrics.}
We evaluate seven proprietary and open-weight models.\footnote{
Gemini 3 Flash~\citep{gemini3flash},
MiniMax-M2.5~\citep{minimax2026m25},
gpt-oss-120b~\citep{openai2025gptoss},
Qwen2.5-72B-Instruct~\citep{qwen25},
Qwen3-Coder-30B-A3B-Instruct~\citep{qwen3coder30b},
Qwen3-Next-80B-A3B-Instruct~\citep{qwen3next80b},
and GLM-4.7~\citep{glm47}.
Figures and tables use shortened names.}
We report Accuracy@$k$ and Recall@$k$ at file, module, and function levels.
Accuracy@$k$ requires all gold targets (up to $k$) to appear in the top-$k$ predictions; Recall@$k$ measures the fraction recovered.
Module-level matching maps predictions to their enclosing module.
Transfer is additionally evaluated on SWE-QA-Bench~\citep{Peng2025SWEQACL}; its setup is described with the RQ3 results.

\subsection{RQ1: File-to-Function Gap on LocBench}
\label{sec:rq1}

\begin{figure}[h]
    \centering

    \begin{subfigure}[b]{\columnwidth}
        \centering
        \includegraphics[width=1\columnwidth]{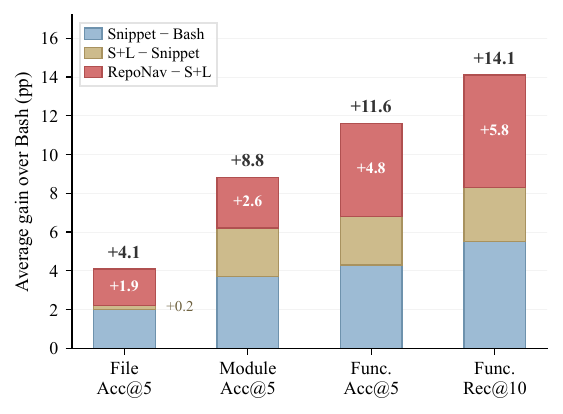}
        \caption{Three-stage gain decomposition.}
        \label{fig:rq1-gain-decomp}
    \end{subfigure}


    \begin{subfigure}[b]{\columnwidth}
        \centering
        \includegraphics[width=1\columnwidth]{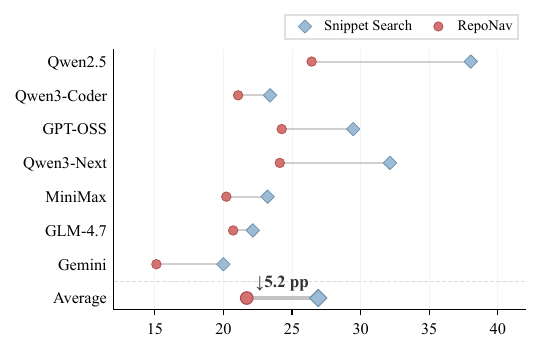}
        \caption{File-to-function gap (pp; lower is better)}
        \label{fig:rq1-gap-dumbbell}
    \end{subfigure}

    \caption{
LocBench gain decomposition and file-to-function gap.
(a) Average gains over Bash are decomposed into retrieval gains (S$-$B), file-structure browsing gains ((S+L)$-$S), and RepoNav gains under matched \texttt{list\_symbols} access (R$-$(S+L)).
(b) RepoNav reduces the average file-to-function gap from 26.9 to 21.7 percentage points compared with Snippet Search.
}
    \label{fig:rq1-bar}
    \vspace{-0.5em}
\end{figure}

Table~\ref{tab:locbench-main} reports the main LocBench results.
To control for tool access, Snippet+ListSym (S+L) adds \texttt{list\_symbols} to the flat snippet interface.
This lets us separate file-structure browsing gains (S$\to$S+L) from RepoNav's additional file-centered organization gains (S+L$\to$R), as visualized in Figure~\ref{fig:rq1-bar}.

\paragraph{Finding 1: File-structure browsing helps, and RepoNav adds further function-level gains.}
Adding \texttt{list\_symbols} to the flat snippet interface (S+L) already improves function-level localization over Snippet Search: average Function Acc@5 rises from 0.484 to 0.509 (+2.5 points).
However, RepoNav yields a further gain to 0.557 (+4.8 points over S+L), indicating that the file-centered scaffold provides substantial additional benefit beyond file-structure browsing access alone.
As Figure~\ref{fig:rq1-gain-decomp} shows, the S+L$\to$R increment is largest at the module and function levels, precisely where within-file navigation matters most.
This two-stage decomposition indicates that RepoNav's gains cannot be fully explained by the availability of \texttt{list\_symbols}; the organization of retrieved evidence into actionable, file-centered cues is a key contributing factor.

\paragraph{Finding 2: The file-to-function gap persists under snippet retrieval; RepoNav narrows it substantially.}
We define the \emph{file-to-function gap} as the difference between File Acc@5 and Function Acc@5.
Under Bash, this gap averages 29.2 points.
Snippet Search slightly narrows it to 26.9 points by improving both file- and function-level accuracy, but the gap remains large because function-level gains do not keep pace with file-level gains, even after relevant files are retrieved.
RepoNav reduces the gap to 21.7 points, a 5.2-point reduction over Snippet Search. This indicates that structured navigation closes a qualitatively different bottleneck than flat retrieval.

\paragraph{Finding 3: RepoNav remains effective even when file-level performance is already high.}
For stronger models such as GLM-4.7 and Gemini-3-Flash, File Acc@5 already exceeds 0.80 under Bash or Snippet Search, yet RepoNav still improves Function Acc@5 by 3.2 and 6.6 points (5.3\% and 10.6\% relative), respectively.
Notably, Snippet Search yields negligible or even slightly negative function-level changes relative to Bash for these models (e.g., $-$0.2 points on Gemini-3-Flash), consistent with the retrieval-only gap reported in Table~\ref{tab:retrieval-only} (Appendix~\ref{app:retrieval}).
RepoNav continues to yield gains, indicating that once relevant files are reachable, the remaining bottleneck increasingly lies in navigating and verifying evidence within those files.
A paired instance-level analysis over all 560 LocBench instances with GPT-OSS-120B further confirms statistically significant improvements over Snippet Search on Function Acc@5 and Function Rec@10 ($p=0.017$ and $p=0.007$, respectively; see Appendix~\ref{app:significance}).
Together, these findings provide evidence that file discovery and function discovery are qualitatively different challenges, and that snippet-style retrieval provides insufficient structural context for the latter.

\subsection{RQ2: Does Structure Help by Volume or by Actionability?}
\label{sec:rq2}

RQ1 shows that RepoNav improves function-level localization, but the mechanism remains unclear.
Does the improvement come from exposing more structural information to the agent, or from organizing retrieved evidence into an action-oriented interface that encourages targeted exploration?

To answer this question, we conduct a controlled ablation on the full 560-instance LocBench benchmark using the same model, dense retrieval backend, search-first workflow, tool access, and verification constraints.
Only the presentation of retrieved evidence differs.
\textbf{File-Only} exposes aggregated candidate files without file-internal structure.
\textbf{Inline Scaffold} exposes richer file-internal structure directly in the retrieval output.
\textbf{Tree Scaffold (RepoNav)} organizes structural evidence into a compact, navigable scaffold with selective cues and explicit next-step hints.
This design separates file-level aggregation, in-context structural volume, and action-oriented organization, allowing us to test whether structure helps by being larger or by being easier to act on.

\begin{table}[htb]
\centering

\small
\setlength{\tabcolsep}{4pt}
\resizebox{\columnwidth}{!}{
\begin{tabular}{@{} l cc cccc @{}}
\toprule
 & \multicolumn{2}{c}{Endpoint Localization} & \multicolumn{4}{c}{Behavior \& Efficiency} \\
\cmidrule(lr){2-3} \cmidrule(l){4-7}
Method & Acc@5 (\%) & Rec@10 (\%) & ListSym\% & Files Insp. & Steps & Tokens \\
\midrule
File-Only       & 32.61 & 44.29 & 39.11 & 2.587 & 10.587 & 50.2k \\
Inline Scaffold & 46.96 & 53.84 & 36.96 & 2.457 &  9.630 & 56.2k \\
Tree Scaffold   & \textbf{52.13} & \textbf{60.98} & \textbf{56.61} & \textbf{2.304} & \textbf{9.391} & \textbf{47.2k} \\
\bottomrule
\end{tabular}%
}
\caption{
RQ2 controlled ablation on the full 560-instance LocBench benchmark using GPT-OSS-120B.
All settings share the same retrieval backend, tool access, workflow, and verification constraints; only evidence presentation differs.
Acc@5/Rec@10 are percentages measuring endpoint localization, while the remaining columns summarize tool use and exploration efficiency.
}
\label{tab:behavior}
\end{table}

\begin{figure}[htb]
    \centering
    \includegraphics[width=\columnwidth]{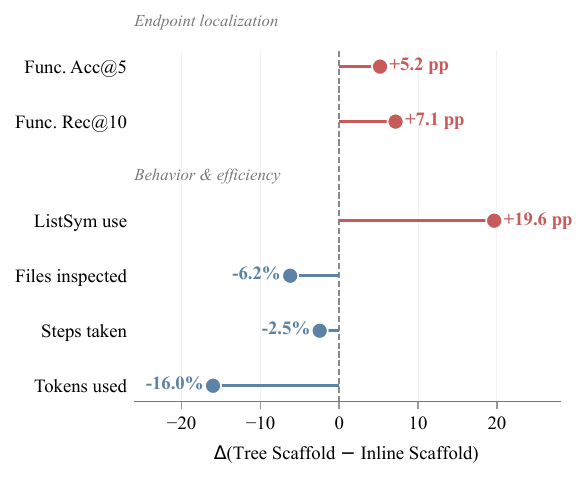}
    \caption{
    Tree Scaffold vs.\ Inline Scaffold in the RQ2 controlled ablation.
    Values report Tree Scaffold minus Inline Scaffold.
    Endpoint localization and ListSym use are absolute percentage-point changes; efficiency metrics are relative percentage changes, where negative values indicate lower cost.
    }
    \label{fig:rq2-delta}
     \vspace{-1.2em}
\end{figure}

Table~\ref{tab:behavior} shows that endpoint localization improves from File-Only to Inline Scaffold and further to Tree Scaffold under matched retrieval, workflow, tool access, and verification constraints.
Function Acc@5 increases from 32.61\% to 46.96\% and 52.13\%, while Function Rec@10 increases from 44.29\% to 53.84\% and 60.98\%.
Thus, exposing file-internal structure is important, but the compact tree scaffold provides additional gains beyond simply placing more structure in the retrieval output.

The behavioral metrics show a similar pattern.
Compared with Inline Scaffold, Tree Scaffold inspects fewer files, takes fewer steps, and uses fewer tokens, while invoking \texttt{list\_symbols} more frequently.
Figure~\ref{fig:rq2-delta} makes this contrast explicit: Tree Scaffold increases \texttt{list\_symbols} use by 19.65 percentage points (56.61\% vs.\ 36.96\%, a 53\% relative increase) while using 16\% fewer tokens.
Together with the evidence-funnel analysis in Appendix~\ref{sec:appendix-funnel}, this demonstrates that Tree Scaffold improves post-inspection actionability rather than simply broadening inspection. A complementary leave-one-out ablation further shows that all three scaffold blocks contribute to function-level localization.
Removing \texttt{[ANCHORS]}, \texttt{[GLIMPSE]}, or \texttt{[CANDIDATE\_TARGETS]} reduces Function Acc@5, with the largest degradation observed when removing \texttt{[CANDIDATE\_TARGETS]} ($-7.6$ points; see Appendix~\ref{app:rq2-ablation}).
The benefit is also concentrated in structurally harder files: under a median-split analysis, RepoNav achieves larger Function Acc@5 gains over Snippet Search for files with more functions (+6.4 points), longer files (+8.2 points), and more sibling symbols (+6.0 points; Appendix~\ref{app:complexity}).

We additionally profile the main interaction costs on GPT-OSS-120B over all 560 LocBench instances.
Compared with Snippet Search, RepoNav uses 62.4k versus 31.7k average trace tokens and 25.3\,s versus 15.5\,s average wall-clock time, while making fewer retrieval-tool calls (1.19 vs.\ 1.39).
RepoNav's scaffold construction itself has a median tool-side latency of 0.34\,s (mean 0.74\,s) and requires no additional repository-wide index or graph.
Accordingly, we use \emph{lightweight} to refer to low incremental infrastructure and deployment overhead, rather than lower interaction-token or latency cost.

\subsection{RQ3: Does the Benefit Transfer Beyond Localization?}
\label{sec:rq3}

The preceding analyses show that RepoNav improves function-level localization on LocBench.
We further ask whether RepoNav also benefits repository-level understanding beyond explicit localization.
To this end, we evaluate on SWE-QA-Bench~\citep{Peng2025SWEQACL}, a repository-level question-answering benchmark that requires agents to locate, inspect, and synthesize evidence from code repositories.
Unlike RQ1, this experiment is intended as a transfer probe of the integrated RepoNav interface rather than a causal ablation of tool access.
We compare three methods, Bash, Snippet Search, and RepoNav, using the same agent setup, interaction budget, and evaluation protocol.
We use the per-model intersection of questions successfully completed and scored by all three methods.

\begin{figure}[htb]
  \centering
  \begin{subfigure}[b]{\columnwidth}
    \includegraphics[width=\columnwidth]{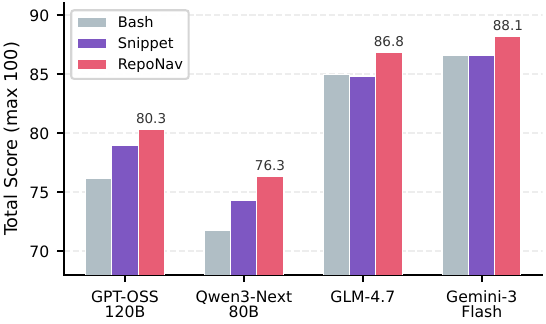}
    \caption{Overall answer quality.}
    \label{fig:rq3_a}
  \end{subfigure}
  \begin{subfigure}[b]{\columnwidth}
    \includegraphics[width=\columnwidth]{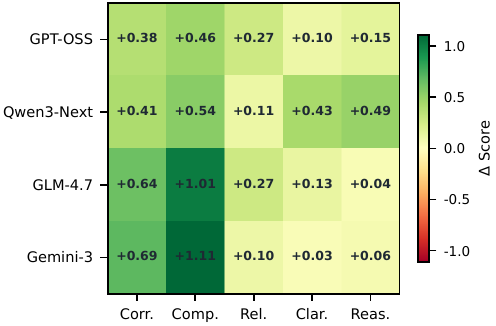}
    \caption{Per-dimension $\Delta$ (RepoNav $-$ Snippet Search).}
    \label{fig:rq3_b}
  \end{subfigure}
  \caption{SWE-QA-Bench overall results.
  (a)~RepoNav achieves the highest total score across
  all four models; exact values in Table~\ref{tab:sweqa-main}.
  (b)~Gains concentrate in Correctness and Completeness,
  consistent with the localization findings.}
  \label{fig:rq3_overall}
\end{figure}

\begin{table}[htb]
\centering

\small
\resizebox{\columnwidth}{!}{%
\begin{tabular}{lccccc}
\toprule
Model & Bash & Snippet & RepoNav & $\Delta_{\mathrm{S}}$ & $\Delta_{\mathrm{B}}$ \\
\midrule
GPT-OSS-120B & 76.12 & 78.97 & \textbf{80.33} & +1.36 & +4.21 \\
Qwen3-Next-80B & 71.72 & 74.31 & \textbf{76.29} & +1.98 & +4.57 \\
GLM-4.7 & 84.96 & 84.75 & \textbf{86.84} & +2.09 & +1.88 \\
Gemini-3-Flash & 86.53 & 86.55 & \textbf{88.14} & +1.59 & +1.61 \\
\midrule
Average & 79.83 & 81.15 & \textbf{82.90} & +1.76 & +3.07 \\
\bottomrule
\end{tabular}%

}
\caption{SWE-QA-Bench total scores.
Scores are reported on the per-model common subset successfully completed and judged for all three methods.
$\Delta_{\mathrm{S}}$ denotes RepoNav minus Snippet Search, and $\Delta_{\mathrm{B}}$ denotes RepoNav minus Bash.
Full dimension-level scores are reported in Appendix~\ref{app:sweqa-full}, Table~\ref{tab:sweqa-full}.}
\label{tab:sweqa-main}
\vspace{-0.5em}
\end{table}

\paragraph{Answer quality.}
Table~\ref{tab:sweqa-main} shows that RepoNav obtains the highest total score across all four models, with an average gain of +1.76 over Snippet Search and +3.07 over Bash.
The gains concentrate in Correctness and Completeness (Figure~\ref{fig:rq3_b}), showing that structured navigation mainly helps agents find and synthesize the right repository evidence rather than merely improving surface-level answer style.
A similar pattern appears for the two strongest models: Snippet Search yields negligible changes relative to Bash, including a slight drop for GLM-4.7 (84.75 vs.\ 84.96) and a near-tie for Gemini-3-Flash (86.55 vs.\ 86.53), whereas RepoNav yields clear improvements (86.84 and 88.14).
A question-type breakdown (Appendix~\ref{sec:appendix-qtype}) shows that gains span all categories, with the largest improvements on \emph{Why} questions requiring cross-file evidence synthesis.
This pattern shows that the scaffold supports repository-level question answering by helping agents assemble evidence chains across functions and files rather than only localizing isolated targets.

\paragraph{Summary.}
The SWE-QA results show that RepoNav also improves repository-level question answering beyond explicit localization.
Across four models, RepoNav consistently outperforms both Bash and Snippet Search.

\section{Conclusion}
This paper investigates how the organization of retrieval outputs shapes the exploration behavior of LLM-based code agents at the repository scale.
Across seven models on LocBench, we identify a persistent file-to-function gap: agents can often reach relevant files, yet still fail to localize the correct function.
RepoNav narrows this gap by reorganizing retrieved snippets into a lightweight, file-centered navigation scaffold without changing the underlying retriever.
Controlled ablations further show that simply exposing more file-structure information is insufficient; RepoNav's gains come from organizing retrieved evidence into a structured, navigable form.
Overall, these findings show that retrieval-augmented code agents depend not only on what evidence is retrieved, but also on how that evidence is organized for exploration.

\section*{Limitations}
Our current evaluation focuses on Python repositories and two benchmarks, LocBench and SWE-QA-Bench.
Although RepoNav is designed as a lightweight interface layer rather than a Python-specific method, our implementation currently uses Python's ast module for file-structure extraction.
We therefore leave validation on repositories written in other programming languages, such as Java, C, C++, Go, and Rust, to future work.
These languages may require language-specific structure extractors, scaffold serialization rules, and evaluation settings.
Extending RepoNav to these settings is an important step toward building a more general and practical repository navigation interface for code agents.
Our experiments focus on localization and repository-level question answering.
Evaluating RepoNav in downstream patch generation, long-horizon maintenance tasks, and interactive developer workflows remains future work.

\bibliography{latex/custom}

\appendix

\section{Parameter Sensitivity Analysis}
\label{sec:appendix-sensitivity}

The hybrid scoring rule uses a weight $\alpha$ to balance peak evidence from the highest-scoring chunk ($s_{f,(1)}$) against consistency across multiple retrieved chunks.
To verify that our results are not sensitive to this choice, we perform a lightweight offline sensitivity analysis: for each of 3 repository-level splits, we reuse the cached dense index and chunk-level scores, recompute file-level rankings for $\alpha \in \{0.0, 0.1, \dots, 1.0\}$, and re-evaluate file retrieval metrics without rerunning the downstream agent.

\begin{table}[ht]
\centering
\small

\setlength{\tabcolsep}{4.5pt}
\begin{tabular}{@{} l cccc @{}}
\toprule
$\alpha$ & Acc@1 & Hit@10 & Recall@10 & MRR \\
\midrule
0.0 {\scriptsize (Avg)} & 0.4965 & 0.8056 & 0.7648 & 0.6077 \\
0.1 & 0.5035 & 0.8056 & 0.7648 & 0.6125 \\
0.2 & 0.5069 & 0.8056 & 0.7648 & 0.6136 \\
0.3 & 0.5208 & 0.8056 & 0.7648 & 0.6231 \\
0.4 & 0.5174 & 0.8056 & 0.7648 & 0.6222 \\
0.5 & \textbf{0.5243} & 0.8056 & 0.7648 & \textbf{0.6251} \\
0.6 & 0.5174 & 0.8090 & 0.7671 & 0.6229 \\
0.7 & 0.5139 & \textbf{0.8125} & \textbf{0.7679} & 0.6212 \\
0.8 & 0.5069 & \textbf{0.8125} & \textbf{0.7679} & 0.6155 \\
0.9 & 0.5104 & \textbf{0.8125} & \textbf{0.7679} & 0.6168 \\
1.0 {\scriptsize (Max)} & 0.5139 & \textbf{0.8125} & \textbf{0.7679} & 0.6168 \\
\bottomrule
\end{tabular}
\caption{File-level retrieval metrics as a function of $\alpha$, aggregated over 3 repository-level development splits.
Performance is stable across the full range: Acc@1 and MRR peak near $\alpha{=}0.5$, while Hit@10 and Recall@10 plateau for $\alpha \ge 0.7$.
We use $\alpha{=}0.5$ in all reported experiments.}
\label{tab:radar-alpha-dev}
\end{table}

As shown in Table~\ref{tab:radar-alpha-dev}, all metrics vary within a narrow band across the full $\alpha$ range.
Ranking-oriented metrics (Acc@1, MRR) peak near $\alpha{=}0.5$, while recall metrics reach a broad plateau for $\alpha \ge 0.7$.
The two extremes, pure averaging ($\alpha{=}0$) and pure max pooling ($\alpha{=}1$), are both competitive, demonstrating that the hybrid formulation is robust rather than requiring careful tuning.
We fix $\alpha{=}0.5$ throughout all experiments reported in this paper.

\section{Retrieval Configuration and Baseline Performance}
\label{app:retrieval}
\label{app:retrieval-details}

Our dense index follows the function-level chunking protocol of \citet{locagent}: each top-level function or method definition is treated as a single chunk and embedded using CodeRankEmbed~\citep{CoRNStack}.
During the agent loop, the agent issues free-form natural language search queries; the retrieval backend returns the top-ranked chunks by cosine similarity.
RepoNav's file-level aggregation rule and the snippet baseline operate on the same retrieved chunks; only the post-retrieval presentation differs.

\paragraph{Raw-chunk retrieval depth.}
The main experiments retrieve 80 raw chunks per query.
To assess sensitivity to this choice, we conduct a retrieval-only analysis over all 560 LocBench instances, varying the raw retrieval depth over $\{20,50,80,100\}$.
We keep the post-aggregation file budget fixed at 15 and use the same CodeRankEmbed index and file-aggregation rule throughout.
Table~\ref{tab:retrieval-depth} reports the resulting retrieval performance.

\begin{table}[ht]
\centering

\small
\setlength{\tabcolsep}{3.5pt}
\begin{tabular}{@{}ccccc@{}}
\toprule
& \multicolumn{2}{c}{Snippet} & \multicolumn{2}{c}{RepoNav} \\
\cmidrule(lr){2-3}\cmidrule(lr){4-5}
Depth & Hit@15 & Rec@15 & Hit@15 & Rec@15 \\
\midrule
20  & 79.64 & 75.28 & 79.46 & 75.01 \\
50  & 81.61 & 78.32 & 81.07 & 77.67 \\
80  & 82.50 & 79.59 & 82.14 & 79.11 \\
100 & 82.86 & 80.04 & 82.32 & 79.34 \\
\bottomrule
\end{tabular}
\caption{
Retrieval-depth sensitivity on all 560 LocBench instances.
The post-aggregation file budget is fixed at 15; all values are percentages.
}
\label{tab:retrieval-depth}
\end{table}

Both retrieval settings exhibit a similar saturation pattern.
Increasing the depth from 20 to 80 yields clear gains, whereas increasing it from 80 to 100 provides only marginal additional improvement.
The default depth of 80 therefore lies near the observed saturation region while avoiding an unnecessarily deeper retrieval pool.

Table~\ref{tab:retrieval-only} reports the retrieval-only accuracy of the shared dense index, evaluated without any downstream agent.
These numbers provide a retrieval-only reference point for the fixed dense index before downstream agent interaction.
The file-to-function localization gap is already visible at the retrieval-only stage:
File Acc@5 reaches 0.721, whereas Function Acc@5 is only 0.348, which is less than half.
Since Snippet Search and RepoNav use the same fixed dense index and raw chunk-level scores, this result supports our interpretation that RepoNav's downstream gains arise from post-retrieval evidence organization and navigation rather than from changes to the underlying retriever.

\begin{table}[ht]
\centering

\resizebox{\columnwidth}{!}{%
\begin{tabular}{l ccc ccc cc}
\toprule
& \multicolumn{3}{c}{File} & \multicolumn{3}{c}{Module} & \multicolumn{2}{c}{Function} \\
\cmidrule(lr){2-4} \cmidrule(lr){5-7} \cmidrule(lr){8-9}
& @1 & @3 & @5 & @5 & @10 & @15 & @5 & @10 \\
\midrule
Acc & 0.505 & 0.661 & 0.721 & 0.521 & 0.604 & 0.666 & 0.348 & 0.430 \\
\bottomrule
\end{tabular}%
}
\caption{Retrieval-only performance of the dense index used by all agent experiments. No downstream agent is involved; scores reflect pure embedding-based retrieval.}
\label{tab:retrieval-only}
\end{table}

\paragraph{Alternative embedder.}
To examine whether RepoNav's benefit transfers beyond the primary CodeRankEmbed retriever, we conduct a fixed 100-instance pilot using UniXcoder (\texttt{microsoft/unixcoder-base}) while retaining the same function-level chunking scheme.
Snippet Search and RepoNav use the same UniXcoder index, GPT-OSS-120B backbone, agent loop, and interaction budget.
Table~\ref{tab:unixcoder} reports the results.

\begin{table}[ht]
\centering

\resizebox{\columnwidth}{!}{%
\begin{tabular}{lcccc}
\toprule
Method & File Hit@5 & Func Hit@5 & Func Acc@5 & Func Rec@10 \\
\midrule
Snippet Search & 76.0 & 60.0 & 19.0 & 36.2 \\
RepoNav        & 79.0 & 63.0 & 23.0 & 41.8 \\
\midrule
$\Delta$       & +3.0 & +3.0 & +4.0 & +5.6 \\
\bottomrule
\end{tabular}%
}
\caption{
Alternative-embedder pilot on a fixed 100-instance LocBench subset using UniXcoder.
All values are percentages.
}
\label{tab:unixcoder}
\end{table}

RepoNav retains positive gains under UniXcoder, improving Function Acc@5 by 4.0 points and Function Rec@10 by 5.6 points over Snippet Search.
This pilot provides preliminary evidence that RepoNav is compatible with a second embedding model, while CodeRankEmbed remains the primary retriever evaluated in the full experiments.

\section{Full LocBench Results}
\label{sec:appendix-full}

Table~\ref{tab:full-results} provides the complete LocBench localization results across all seven models, four exploration settings, and the reported Accuracy@$k$ and Recall@$k$ variants.
The main text (Table~\ref{tab:locbench-main}) reports File Acc@5, Module Acc@5, Function Acc@5, and Function Rec@10; this table additionally includes lower- and higher-rank Accuracy@$k$ and Recall@$k$ variants where applicable, enabling fine-grained comparison across the reported $k$ values.

Several patterns emerge from the complete results that are not visible in the summary table.
First, RepoNav's gains are consistent across $k$ values: improvements at Acc@1 tend to be slightly smaller than those at larger $k$, suggesting that RepoNav helps agents recover more of the gold target set within the candidate budget rather than always ranking all required targets first.
Second, the recall-level gains are generally larger than the accuracy-level gains at comparable $k$, reflecting that RepoNav helps agents localize a greater fraction of the total gold targets per instance.
Third, the file-level recall columns show that RepoNav largely preserves file-level coverage while improving module- and function-level localization, although small drops appear for some models.


\begin{table*}[ht]
\centering

\scriptsize
\setlength{\tabcolsep}{3.2pt}
\resizebox{\textwidth}{!}{%
\begin{tabular}{llccc|cccc|cccc|ccc|ccc|ccc}
\toprule
& & \multicolumn{3}{c|}{File Acc} & \multicolumn{4}{c|}{Module Acc} & \multicolumn{4}{c|}{Function Acc} & \multicolumn{3}{c|}{File Recall} & \multicolumn{3}{c|}{Module Recall} & \multicolumn{3}{c}{Function Recall} \\
Model & Setting & @1 & @3 & @5 & @1 & @3 & @5 & @10 & @1 & @3 & @5 & @10 & @3 & @5 & @10 & @3 & @5 & @10 & @3 & @5 & @10 \\
\midrule
\multirow{4}{*}{Qwen2.5-72B}
& Bash             & 0.4714 & 0.5679 & 0.5893 & 0.3429 & 0.3661 & 0.3964 & 0.4214 & 0.2571 & 0.2339 & 0.2536 & 0.2643 & 0.5929 & 0.6206 & 0.6295 & 0.4048 & 0.4345 & 0.4622 & 0.2756 & 0.2946 & 0.3067 \\
& Snippet Search   & 0.6304 & 0.6607 & 0.6804 & 0.4357 & 0.4446 & 0.4607 & 0.4696 & 0.3196 & 0.2821 & 0.3000 & 0.3089 & 0.6985 & 0.7179 & 0.7223 & 0.4943 & 0.5176 & 0.5272 & 0.3482 & 0.3663 & 0.3771 \\
& Snippet+ListSym  & 0.6464 & 0.6696 & 0.6732 & 0.5229 & 0.5139 & 0.5621 & 0.5386 & 0.3886 & 0.3629 & 0.3843 & 0.4064 & 0.7063 & 0.7106 & 0.7112 & 0.5405 & 0.5785 & 0.5886 & 0.4193 & 0.4436 & 0.4901 \\
& RepoNav          & \textbf{0.6661} & \textbf{0.6911} & \textbf{0.6982} & \textbf{0.5411} & \textbf{0.5375} & \textbf{0.5696} & \textbf{0.5929} & \textbf{0.4071} & \textbf{0.3946} & \textbf{0.4339} & \textbf{0.4786} & \textbf{0.7301} & \textbf{0.7382} & \textbf{0.7382} & \textbf{0.5914} & \textbf{0.6297} & \textbf{0.6488} & \textbf{0.4634} & \textbf{0.5113} & \textbf{0.5492} \\
\midrule
\multirow{4}{*}{GPT-OSS-120B}
& Bash             & 0.6179 & 0.6875 & 0.7054 & 0.4679 & 0.4643 & 0.5000 & 0.5196 & 0.3643 & 0.3214 & 0.3393 & 0.3536 & 0.7223 & 0.7397 & 0.7457 & 0.5167 & 0.5549 & 0.5742 & 0.3798 & 0.4007 & 0.4171 \\
& Snippet Search   & 0.7125 & 0.7518 & 0.7589 & 0.5964 & 0.5786 & 0.6125 & 0.6357 & 0.4929 & 0.4393 & 0.4643 & 0.4893 & 0.7887 & 0.7967 & 0.7976 & 0.6372 & 0.6748 & 0.6936 & 0.5226 & 0.5462 & 0.5707 \\
& Snippet+ListSym  & \textbf{0.7268} & 0.7500 & 0.7589 & 0.5929 & 0.6007 & 0.6304 & 0.6429 & 0.4911 & 0.4521 & 0.4875 & 0.5089 & 0.7863 & 0.7967 & 0.7976 & 0.6567 & 0.6815 & 0.7020 & 0.5409 & 0.5663 & 0.5858 \\
& RepoNav          & 0.7250 & \textbf{0.7589} & \textbf{0.7696} & \textbf{0.5982} & \textbf{0.6071} & \textbf{0.6500} & \textbf{0.6625} & \textbf{0.5064} & \textbf{0.4868} & \textbf{0.5271} & \textbf{0.5486} & \textbf{0.7964} & \textbf{0.8056} & \textbf{0.8074} & \textbf{0.6810} & \textbf{0.7163} & \textbf{0.7269} & \textbf{0.5636} & \textbf{0.5990} & \textbf{0.6184} \\
\midrule
\multirow{4}{*}{Qwen3-Next-80B}
& Bash             & 0.6482 & 0.7107 & 0.7339 & 0.4804 & 0.4929 & 0.5232 & 0.5464 & 0.3696 & 0.3321 & 0.3393 & 0.3500 & 0.7430 & 0.7668 & 0.7760 & 0.5352 & 0.5701 & 0.5936 & 0.3751 & 0.3944 & 0.4098 \\
& Snippet Search   & 0.7179 & 0.7286 & 0.7321 & 0.5536 & 0.5589 & 0.5679 & 0.5804 & 0.4482 & 0.4054 & 0.4107 & 0.4268 & 0.7698 & 0.7764 & 0.7764 & 0.6062 & 0.6346 & 0.6520 & 0.4634 & 0.4957 & 0.5170 \\
& Snippet+ListSym  & 0.7103 & 0.7304 & 0.7329 & 0.5807 & 0.5818 & 0.5886 & 0.6464 & 0.4986 & 0.4557 & 0.4539 & 0.4718 & 0.7380 & 0.7535 & 0.7557 & 0.6200 & 0.6542 & 0.6768 & 0.5010 & 0.5309 & 0.5636 \\
& RepoNav          & \textbf{0.7339} & \textbf{0.7357} & \textbf{0.7393} & \textbf{0.6018} & \textbf{0.6018} & \textbf{0.6321} & \textbf{0.6482} & \textbf{0.5146} & \textbf{0.4893} & \textbf{0.4982} & \textbf{0.5268} & \textbf{0.7770} & \textbf{0.7814} & \textbf{0.7819} & \textbf{0.6562} & \textbf{0.7000} & \textbf{0.7212} & \textbf{0.5355} & \textbf{0.5774} & \textbf{0.6215} \\
\midrule
\multirow{4}{*}{Qwen3-Coder-30B}
& Bash             & 0.6429 & 0.6857 & 0.7000 & 0.5393 & 0.5464 & 0.5643 & 0.5732 & 0.4571 & 0.4089 & 0.4196 & 0.4232 & 0.7215 & 0.7375 & 0.7396 & 0.5834 & 0.6186 & 0.6322 & 0.4523 & 0.4858 & 0.4982 \\
& Snippet Search   & 0.6786 & 0.6893 & 0.6929 & 0.5679 & 0.5571 & 0.5839 & 0.5875 & 0.4804 & 0.4464 & 0.4589 & 0.4679 & 0.7225 & 0.7282 & 0.7309 & 0.6035 & 0.6429 & 0.6537 & 0.4972 & 0.5321 & 0.5507 \\
& Snippet+ListSym  & 0.6804 & 0.6964 & 0.7071 & 0.5857 & 0.5768 & 0.5982 & 0.6036 & 0.4964 & 0.4643 & 0.4757 & 0.4982 & 0.7326 & 0.7436 & 0.7454 & 0.6235 & 0.6577 & 0.6663 & 0.5231 & 0.5607 & 0.5717 \\
& RepoNav          & \textbf{0.6946} & \textbf{0.7196} & \textbf{0.7268} & \textbf{0.6036} & \textbf{0.6036} & \textbf{0.6286} & \textbf{0.6357} & \textbf{0.5054} & \textbf{0.4893} & \textbf{0.5161} & \textbf{0.5393} & \textbf{0.7555} & \textbf{0.7623} & \textbf{0.7623} & \textbf{0.6424} & \textbf{0.6887} & \textbf{0.6997} & \textbf{0.5303} & \textbf{0.5887} & \textbf{0.6180} \\
\midrule
\multirow{4}{*}{MiniMax-M2.5}
& Bash             & 0.7354 & 0.7425 & 0.7604 & 0.6193 & 0.6264 & 0.6604 & 0.6729 & 0.5443 & 0.5086 & 0.5246 & 0.5407 & 0.7771 & 0.7948 & 0.8098 & 0.6669 & 0.7141 & 0.7347 & 0.5525 & 0.6038 & 0.6328 \\
& Snippet Search   & 0.7393 & 0.7518 & 0.7643 & 0.6321 & 0.6429 & 0.6714 & 0.6857 & 0.5500 & 0.5268 & 0.5321 & 0.5482 & 0.7822 & 0.7963 & 0.7996 & 0.6847 & 0.7298 & 0.7485 & 0.5722 & 0.6157 & 0.6409 \\
& Snippet+ListSym  & 0.7589 & 0.7732 & 0.7804 & 0.6482 & 0.6411 & 0.6714 & 0.6893 & 0.5696 & 0.5339 & 0.5339 & 0.5589 & 0.8093 & 0.8199 & 0.8235 & 0.6858 & 0.7345 & 0.7566 & 0.5712 & 0.6229 & 0.6519 \\
& RepoNav          & \textbf{0.7736} & \textbf{0.7843} & \textbf{0.8032} & \textbf{0.6532} & \textbf{0.6586} & \textbf{0.6907} & \textbf{0.6996} & \textbf{0.5882} & \textbf{0.5921} & \textbf{0.6011} & \textbf{0.6136} & \textbf{0.8135} & \textbf{0.8347} & \textbf{0.8471} & \textbf{0.7073} & \textbf{0.7449} & \textbf{0.7696} & \textbf{0.6279} & \textbf{0.6869} & \textbf{0.7084} \\
\midrule
\multirow{4}{*}{GLM-4.7}
& Bash             & 0.7568 & 0.7800 & 0.8079 & 0.6496 & 0.6425 & 0.6854 & 0.7264 & 0.5782 & 0.5639 & 0.5854 & 0.6300 & 0.8031 & 0.8207 & 0.8260 & 0.6840 & 0.7324 & 0.7763 & 0.5987 & 0.6556 & 0.7137 \\
& Snippet Search   & 0.7875 & 0.8036 & 0.8214 & 0.6768 & 0.6625 & 0.6982 & 0.7268 & 0.6054 & 0.5875 & 0.6000 & 0.6357 & 0.8405 & 0.8554 & 0.8634 & 0.7044 & 0.7595 & 0.7907 & 0.6224 & 0.6765 & 0.7223 \\
& Snippet+ListSym  & 0.7857 & 0.8036 & 0.8125 & 0.6607 & 0.6786 & 0.7089 & 0.7304 & 0.5804 & 0.5607 & 0.5768 & 0.6036 & 0.8396 & 0.8470 & 0.8506 & 0.7197 & 0.7665 & 0.7896 & 0.6007 & 0.6582 & 0.6950 \\
& RepoNav          & \textbf{0.7886} & \textbf{0.8064} & \textbf{0.8389} & \textbf{0.6950} & \textbf{0.6968} & \textbf{0.7336} & \textbf{0.7586} & \textbf{0.6343} & \textbf{0.5950} & \textbf{0.6318} & \textbf{0.6771} & \textbf{0.8525} & \textbf{0.8765} & \textbf{0.8889} & \textbf{0.7683} & \textbf{0.8144} & \textbf{0.8200} & \textbf{0.6681} & \textbf{0.7509} & \textbf{0.7826} \\
\midrule
\multirow{4}{*}{Gemini-3-Flash}
& Bash             & 0.8089 & \textbf{0.8250} & 0.8357 & 0.7036 & 0.6982 & 0.7089 & 0.7196 & 0.6286 & 0.6054 & 0.6250 & 0.6464 & \textbf{0.8637} & \textbf{0.8743} & \textbf{0.8761} & 0.7434 & 0.7731 & 0.7868 & 0.6558 & 0.7041 & 0.7365 \\
& Snippet Search   & 0.8161 & 0.8143 & 0.8232 & 0.7089 & 0.6857 & 0.7018 & 0.7018 & 0.6357 & 0.6089 & 0.6232 & 0.6304 & 0.8526 & 0.8612 & 0.8612 & 0.7337 & 0.7654 & 0.7712 & 0.6558 & 0.7028 & 0.7223 \\
& Snippet+ListSym  & 0.8078 & 0.8115 & 0.8221 & 0.7154 & 0.7026 & 0.7165 & 0.7242 & 0.6363 & 0.6327 & 0.6481 & 0.6519 & 0.8372 & 0.8465 & 0.8583 & 0.7296 & 0.7702 & 0.8015 & 0.6624 & 0.7151 & 0.7387 \\
& RepoNav          & \textbf{0.8196} & 0.8207 & \textbf{0.8404} & \textbf{0.7536} & \textbf{0.7393} & \textbf{0.7518} & \textbf{0.7625} & \textbf{0.6857} & \textbf{0.6750} & \textbf{0.6893} & \textbf{0.7125} & 0.8603 & 0.8691 & 0.8700 & \textbf{0.7860} & \textbf{0.8140} & \textbf{0.8252} & \textbf{0.7105} & \textbf{0.7666} & \textbf{0.8026} \\
\bottomrule
\end{tabular}%
}
\caption{Complete LocBench localization results with all @$k$ variants. This table supplements Table~\ref{tab:locbench-main} with additional Acc@$k$ and Recall@$k$ values not shown in the main text. Bold indicates the best setting within each model block.}
\label{tab:full-results}
\end{table*}

\subsection{Statistical Significance}
\label{app:significance}

All configurations use greedy decoding ($\mathrm{temperature}=0$), and each configuration is evaluated once on every instance.
We conduct a paired instance-level analysis over all 560 LocBench instances using GPT-OSS-120B to quantify uncertainty in the primary function-level results.

\begin{table}[t]
\centering
\small

\setlength{\tabcolsep}{4pt}
\begin{tabular}{@{}lccc@{}}
\toprule
Metric & $\Delta$ & 95\% CI & $p$-value \\
\midrule
Function Acc@5 & +6.28 pp & $[+0.7,+6.4]$ & 0.017 \\
Function Rec@10 & +4.77 pp & $[+1.0,+6.2]$ & 0.007 \\
\bottomrule
\end{tabular}
\caption{Paired instance-level analysis of RepoNav versus Snippet Search over all 560 LocBench instances with GPT-OSS-120B. Confidence intervals are 95\% paired-bootstrap CIs.}
\label{tab:significance}
\end{table}

Both confidence intervals exclude zero, providing statistical evidence for the improvements on the two primary function-level localization metrics.

\section{Additional RQ2 Analyses}
\label{sec:appendix-rq2}

\subsection{Evidence Funnel}
\label{sec:appendix-funnel}

To complement the behavioral analysis in Section~\ref{sec:rq2}, we report three diagnostic trajectory statistics:
whether the gold file becomes available to the agent either through the retrieved candidate set or through subsequent shell exploration (\emph{Coverage}),
whether the agent inspects it given coverage (\emph{Inspection}),
and whether inspection leads to correct localization (\emph{Resolution}).
These statistics are coarse trajectory diagnostics rather than mutually exhaustive paths through the agent workflow, and should not be interpreted as a multiplicative decomposition of the endpoint localization metrics in Table~\ref{tab:behavior}.

\begin{table}[ht]
\centering

\small
\begin{tabular}{lccc}
\toprule
Setting & Coverage & Inspection & Resolution \\
\midrule
File-Only       & 78.26\% & 97.22\% & 92.50\% \\
Inline Scaffold & 80.43\% & \textbf{97.30\%} & 92.68\% \\
Tree Scaffold   & \textbf{82.61\%} & 92.11\% & \textbf{97.50\%} \\
\bottomrule
\end{tabular}
\caption{
Evidence funnel across the three RQ2 interface variants.
Each stage conditions on the previous one and should be read as a diagnostic decomposition, not as a multiplicative estimate of endpoint localization performance.
All three settings share the same dense retrieval backend; Coverage can still vary because agents may discover files through shell exploration after the initial retrieval output.
}
\label{tab:funnel}
\end{table}

As Table~\ref{tab:funnel} shows, Inline Scaffold and Tree Scaffold exhibit different diagnostic profiles.
Inline Scaffold attains the highest inspection rate (97.30\%), consistent with an information-heavy interface that encourages the agent to read broadly.
Tree Scaffold leads to more selective inspection (92.11\%) but achieves the highest resolution rate once the gold file is inspected (97.50\%), consistent with the more frequent \texttt{list\_symbols} usage and fewer wasted steps reported in Table~\ref{tab:behavior}.

The key difference between these two behavioral profiles can be summarized as follows.
Inline Scaffold maximizes the probability of \emph{looking at} the gold file: by embedding rich structural detail directly in the retrieval output, it lowers the cost of passive browsing.
Tree Scaffold instead maximizes the probability of \emph{correctly acting on} the gold file once inspected: by presenting compact cues with explicit next-step prompts, it encourages the agent to actively verify candidates through tool use rather than relying on in-context information alone.
This decomposition reinforces the conclusion that Tree Scaffold's advantage lies in the quality of post-inspection navigation rather than the breadth of initial coverage.

\subsection{Block-level Scaffold Ablation}
\label{app:rq2-ablation}

To complement the representation-level comparison in Section~\ref{sec:rq2}, we conduct a matched leave-one-out ablation of the three RepoNav scaffold blocks on all 560 LocBench instances using GPT-OSS-120B.
All variants use identical prompts, tools, interaction budgets, metrics, and evaluation data, with one scaffold block removed at a time.

\begin{table}[ht]
\centering
\small
\setlength{\tabcolsep}{3pt}
\begin{tabular}{@{}lcccc@{}}
\toprule
Variant & Acc@5 & Rec@10 & ListSym\% & Tokens \\
\midrule
Full Scaffold
& \textbf{52.13}
& \textbf{60.98}
& 56.61
& \textbf{52.6k} \\
w/o Anchors
& 48.90
& 57.10
& 68.00
& 62.8k \\
w/o Glimpse
& 49.60
& 58.10
& 66.20
& 58.7k \\
w/o Targets
& 44.50
& 53.50
& 36.60
& 54.2k \\
\bottomrule
\end{tabular}
\caption{
Block-level leave-one-out ablation of the RepoNav scaffold.
Acc@5 and Rec@10 denote function-level localization performance.
}
\label{tab:block-ablation}
\end{table}

Removing any scaffold block reduces function-level localization performance.
Removing \texttt{[ANCHORS]} decreases Function Acc@5 by 3.2 points and Function Rec@10 by 3.9 points, while removing \texttt{[GLIMPSE]} decreases them by 2.5 and 2.9 points, respectively.
The largest degradation occurs without \texttt{[CANDIDATE\_TARGETS]}, where Function Acc@5 drops by 7.6 points and Function Rec@10 by 7.5 points.
Moreover, removing \texttt{[ANCHORS]} or \texttt{[GLIMPSE]} increases both \texttt{list\_symbols} use and token consumption, suggesting that these blocks reduce additional manual browsing.
Overall, the full scaffold achieves the strongest accuracy--token trade-off among the evaluated variants.

\subsection{Complexity-Stratified Analysis}
\label{app:complexity}

To examine whether RepoNav's benefit is associated with within-file difficulty, we perform a median-split analysis over all 560 LocBench instances using GPT-OSS-120B.
We stratify instances according to three properties of the gold file: number of functions, file length, and number of sibling symbols.

\begin{table}[ht]
\centering

\small
\setlength{\tabcolsep}{5pt}
\begin{tabular}{lcc}
\toprule
Higher-complexity subset & $\Delta$ Acc@5 & $\Delta$ Rec@10 \\
\midrule
More functions        & +6.4 & +6.1 \\
Longer files          & +8.2 & +7.3 \\
More sibling symbols  & +6.0 & +5.9 \\
\bottomrule
\end{tabular}
\caption{
RepoNav gains over Snippet Search on the higher-complexity half of LocBench under three median-split criteria.
Values are absolute percentage-point improvements.
}
\label{tab:complexity}
\end{table}

RepoNav's gains are consistently larger on files with greater within-file complexity, while the corresponding gains on simpler files are small.
This pattern supports the interpretation that RepoNav primarily helps agents discriminate among plausible targets after reaching a relevant file, rather than merely improving initial file discovery.

\section{SWE-QA Full Results}
\label{app:sweqa-full}

\subsection{Dimension-Level Scores}

Table~\ref{tab:sweqa-full} presents the full dimension-level SWE-QA-Bench results, complementing the aggregate total scores reported in Table~\ref{tab:sweqa-main}.
Each answer is independently scored on a 20-point scale across five dimensions: Correctness, Completeness, Relevance, Clarity, and Reasoning.

Across all four models, RepoNav achieves the highest score on every individual dimension.
The gains are most pronounced in Correctness and Completeness, consistent with the hypothesis that structured navigation helps agents find and synthesize the right evidence rather than merely improving surface-level answer quality.
Relevance, Clarity, and Reasoning show smaller but consistently positive improvements, indicating that better evidence navigation has a downstream effect on overall answer coherence.

\begin{table}[ht]
\centering

\small
\resizebox{\columnwidth}{!}{%
\begin{tabular}{llcccccc}
\toprule
Model & Method & Corr. & Comp. & Rel. & Clar. & Reas. & Total \\
\midrule
\multirow{3}{*}{GPT-OSS-120B}
& Bash     & 14.06 & 11.68 & 17.55 & 16.54 & 16.29 & 76.12 \\
& Snippet Search & 14.46 & 13.20 & 18.08 & 16.72 & 16.51 & 78.97 \\
& RepoNav  & \textbf{14.84} & \textbf{13.66} & \textbf{18.35} & \textbf{16.82} & \textbf{16.66} & \textbf{80.33} \\
\midrule
\multirow{3}{*}{Qwen3-Next-80B}
& Bash     & 12.44 & 10.52 & 17.91 & 15.71 & 15.14 & 71.72 \\
& Snippet Search & 13.22 & 11.78 & 17.96 & 15.88 & 15.47 & 74.31 \\
& RepoNav  & \textbf{13.63} & \textbf{12.32} & \textbf{18.07} & \textbf{16.31} & \textbf{15.96} & \textbf{76.29} \\
\midrule
\multirow{3}{*}{GLM-4.7}
& Bash     & 16.11 & 15.96 & 17.93 & 17.50 & 17.46 & 84.96 \\
& Snippet Search & 16.05 & 15.96 & 17.81 & 17.50 & 17.43 & 84.75 \\
& RepoNav  & \textbf{16.69} & \textbf{16.97} & \textbf{18.08} & \textbf{17.63} & \textbf{17.47} & \textbf{86.84} \\
\midrule
\multirow{3}{*}{Gemini-3-Flash}
& Bash     & 16.39 & 16.16 & 18.17 & 17.91 & 17.86 & 86.53 \\
& Snippet Search & 16.35 & 16.02 & 18.11 & 17.90 & 17.87 & 86.55 \\
& RepoNav  & \textbf{17.04} & \textbf{17.13} & \textbf{18.21} & \textbf{17.93} & \textbf{17.93} & \textbf{88.14} \\
\bottomrule
\end{tabular}%
}
\caption{SWE-QA-Bench dimension-level results.
Results are reported on the per-model common subset successfully scored for all three methods.
Each dimension is scored on a 20-point scale; Total is their sum (max 100).
Dimensions: Correctness (Corr.), Completeness (Comp.), Relevance (Rel.), Clarity (Clar.), and Reasoning (Reas.).}
\label{tab:sweqa-full}
\end{table}

\subsection{Question-Type Breakdown}
\label{sec:appendix-qtype}

\begin{figure*}[ht]
  \centering
  \begin{subfigure}[b]{0.48\textwidth}
    \includegraphics[width=\textwidth]{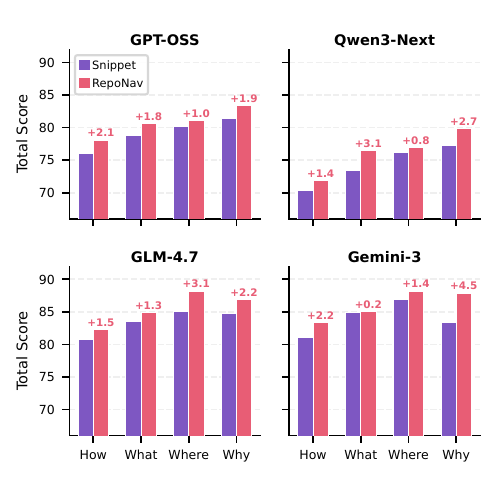}
    \caption{Score by question type.}
    \label{fig:rq3_score}
  \end{subfigure}
  \hfill
  \begin{subfigure}[b]{0.48\textwidth}
    \includegraphics[width=\textwidth]{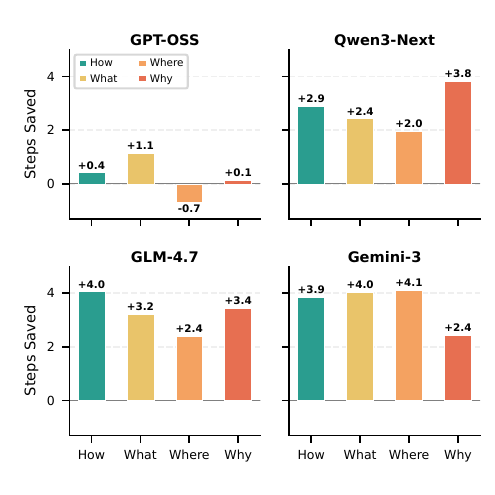}
    \caption{Interaction savings by question type.}
    \label{fig:rq3_steps}
  \end{subfigure}
  \caption{SWE-QA-Bench question-type breakdown across four models.
  (a)~Score improvements span all question types, with the largest gains on \emph{Why} questions.
  (b)~Interaction savings are consistent for three of four models; GPT-OSS-120B shows mixed results.
  Model names are abbreviated for space; see Section~\ref{sec:setup} for full names.}
  \label{fig:rq3_qtype}
\end{figure*}

Figure~\ref{fig:rq3_qtype} breaks down SWE-QA-Bench performance by question type (\emph{How}, \emph{What}, \emph{Where}, \emph{Why}) across all four evaluated models.
The analysis complements the aggregate results in Table~\ref{tab:sweqa-main} by showing that RepoNav's gains are not confined to a single question category.

Score improvements are observed across all types, with the largest and most consistent gains on \emph{Why} questions.
This is consistent with the localization findings: \emph{Why} questions typically require synthesizing evidence across multiple files and tracing causal chains through the codebase---precisely the scenario where structured navigation helps agents avoid premature anchoring on a single entry point.

Interaction savings (measured in steps) are substantial for three of four models.
GPT-OSS-120B shows smaller and mixed savings, suggesting that efficiency gains may depend on model-specific exploration behavior rather than model strength alone.
The explicit continuation cues in RepoNav therefore appear to interact with each model's default exploration policy.

\section{Additional Case Studies}
\label{sec:appendix-cases}

We provide additional analyses illustrating navigation regimes and residual failure modes beyond those highlighted in the main text.
Figure~\ref{fig:motivating_example} illustrates the core same-file disambiguation mechanism; below we describe a cross-file pivoting case, summarize residual file-to-function failures, and present a representative remaining failure boundary.

\smallskip
\noindent\textbf{Cross-file pivoting (LocBench).}\par
\noindent
In \texttt{DS4SD\_\_docling-314}, the visible symptom appears near Markdown output code, while the gold target lies in an upstream implementation file, \texttt{msword\_backend.py}.
Snippet-style search keeps the agent near the output-side sink, since the highest-scoring chunks come from Markdown-related code that shares surface keywords with the issue.
RepoNav retains the upstream implementation file in its candidate structure because its hybrid file-level scoring rule rewards files supported by multiple moderately scored chunks, rather than only by a single dominant match.
The tree scaffold then exposes both symptom-side and source-side files in a navigable list, enabling the agent to pivot from output-related code to the gold implementation target.

\smallskip
\noindent\textbf{Residual failure diagnostic (LocBench).}\par
\noindent
We further analyze the 69 GPT-OSS-120B cases in which RepoNav reaches the gold file but misses the target function.
An automatic diagnostic based on rule-based file-complexity and trajectory signals attributes 71.0\% of these failures to ambiguous or similar sibling functions, 17.4\% to long or crowded gold files, and 11.6\% to within-file ranking or incomplete verification, indicating that most residual errors arise from fine-grained discrimination within the correct file.

\smallskip
\noindent\textbf{Remaining failure boundary (LocBench).}\par
\noindent
In \texttt{yt-dlp\_\_yt-dlp-11615}, RepoNav successfully brings the agent to the correct file, but the agent still stops at higher-level wrapper methods rather than drilling down to the gold targets, which are lower-level helper functions called by the wrapper.
The \texttt{[GLIMPSE]} block lists these helpers, but the agent does not inspect them further, treating the wrapper as a sufficient answer.
This suggests that RepoNav can reduce wrong-file fixation, but deep same-file evidence harvesting remains an open problem, particularly when the gold target is multiple call-hops away from the most salient entry point.

\section{RepoNav Scaffold Specification and Output Format}
\label{app:scaffold-spec}
\label{sec:appendix-output}

\subsection{Serialization Budgets and Ordering}

RepoNav uses a single pre-specified set of serialization budgets and deterministic ordering rules across all experiments to keep the scaffold compact and reproducible.
For each candidate file, \texttt{[ANCHORS]} includes at most two symbols.
Anchors are selected by case-insensitive substring matching between query tokens and symbol names, and ranked by token overlap, symbol-kind priority, and span length.

\texttt{[GLIMPSE]} includes up to three non-anchor symbols per file, selected by query-aware ranking while preserving symbol-kind diversity.
If the top-ranked file has no anchor match, this budget is increased to five to expose a broader file sketch.

\texttt{[CANDIDATE\_TARGETS]} is capped at four entries.
Candidates are ordered first by anchors, then by same-file call-neighborhood symbols of anchors, and finally by selected glimpse symbols.
A fixed continuation cue is appended whenever a candidate block is present.

Call context is name-only: it records same-file caller and callee symbol names without function bodies, arguments, or interprocedural analysis.
By default, RepoNav parses the top five candidate files and auto-expands full three-block scaffolds for the top three.

\subsection{Serialized Output Example}

\begin{figure}[ht]
\begin{lstlisting}[basicstyle=\ttfamily\scriptsize, frame=single, breaklines=true, numbers=none]
[DIR] src/backend/
  [FILE] backend/server.py (evidence: 3)
     |-- imports-by <- backend/app.py
     |-- [ANCHORS]
     |   `-- oauth_callback(function)[L120-L156]
     |         invokes -> validate_token, init_app
     |-- [GLIMPSE]
     |   `-- class HTTPServer:
     |   `-- def handle_request():
     `-- [CANDIDATE_TARGETS]
         - backend/server.py:oauth_callback
         - backend/server.py:validate_token
         >> Next: run `list_symbols` on this file to
            inspect sibling symbols.
\end{lstlisting}
\caption{A truncated example of RepoNav's serialized output.
The indentation-based tree provides file-centered organization, a compact structural sketch, and explicit continuation cues.
The plain-text format requires no specialized query language from the agent.}
\label{fig:reponav_output}
\end{figure}

Figure~\ref{fig:reponav_output} shows a truncated example of RepoNav's serialized output as presented to the agent.
The indentation-based tree provides file-centered organization without requiring any specialized query language or structured API from the agent.

The three blocks---\texttt{[ANCHORS]}, \texttt{[GLIMPSE]}, and \texttt{[CANDIDATE\_TARGETS]}---correspond to the design principles described in Section~\ref{sec:design-principles}.
The anchor block provides grounded entry points with call-context annotations (\texttt{invokes ->} and \texttt{invoked-by <-}).
The glimpse block exposes non-anchor symbols as a structural sketch, preventing fixation on anchors alone.
The candidate targets block aggregates actionable options and ends with an explicit continuation cue
(\texttt{>> Next: run \textasciigrave list\_symbols\textasciigrave\ on this file to inspect sibling symbols}),
lowering the cost of continued exploration.

\end{document}